\documentclass[showpacs,aps,twocolumn,floatfix]{revtex4}
\usepackage{graphicx,epsfig}
\usepackage{times}
\usepackage{amsmath,amssymb}

\newlength\figurewidth
\def\R{\hspace{-1.mm}\mbox{\makebox[.2em][l]{I}R}}

\newcommand{\fig}{Fig.}

\newcommand{\rem}[1]{}

\newcommand{\imag}[1]{\text{Im}(#1)}

\newcommand{\real}[1]{\text{Re}(#1)}

\newcommand{\neff}{n}

\begin{document}

\title{Fractal Weyl law for chaotic microcavities:
  Fresnel's laws imply multifractal scattering}
\author{Jan Wiersig}
\affiliation{Institut f{\"u}r Theoretische Physik, Universit{\"a}t Bremen,
  Postfach 330 440, 28334 Bremen, Germany}
\author{J{\"o}rg Main}
\affiliation{1. Institut f{\"u}r Theoretische Physik, Universit{\"a}t
  Stuttgart, 70550 Stuttgart, Germany}
\date{\today}

\begin{abstract}
We demonstrate that the harmonic inversion technique is a powerful tool to
analyze the spectral properties of optical microcavities. As an interesting 
example we study the statistical properties of complex frequencies of the
fully chaotic microstadium. We show that the conjectured fractal Weyl
law for open chaotic systems [W. T. Lu, S. Sridhar, and M. Zworski,
Phys. Rev. Lett. {\bf 91}, 154101 (2003)] is valid for dielectric
microcavities only if the concept of the chaotic repeller is extended
to a multifractal by incorporating Fresnel's laws.

\end{abstract}
% 42.25.-p Wave optics, includes:
%          42.25.Gy Edge and boundary effects; reflection and refraction
%          42.25.Fx Diffraction and scattering
%          42.25.Hz Interference
% 05.45.Mt Quantum chaos; semiclassical methods
\maketitle

% introduction
% Optical microcavities
\section{Introduction}
Optical microcavities are expected to be key components in future photonic 
applications, such as ultralow threshold lasers~\cite{Park04,Ulrich06}, 
single-photon emitters~\cite{KBKY99,Michler2000}, and correlated photon-pair
emitters~\cite{BUMWJF04b}. Microdisk cavities with
non-circular cross-sectional shape have recently attracted considerable
interest in the quantum chaos community since the internal ray dynamics is 
non-integrable~\cite{ND97,GCNNSFSC98}. In fact, microdisks can be
considered as {\it open billiards}. In a billiard a point-like particle moves
freely in a two-dimensional domain with elastic reflections at the
boundary~\cite{Berry81}.  
Depending on the shape of the boundary the system shows a variety of
dynamical behaviors ranging from integrable to fully chaotic~\cite{Robnik83}. 
Light rays in a microcavity behave similarly as they are totally reflected at
the boundary as long as the angle of incidence $\chi$ (measured from the
normal) is larger than the
critical angle for total internal reflection $\chi_\text{c} = \arcsin{(1/n)}$,
where $n$ is the index of refraction. If,
however, $\chi < \chi_\text{c}$ then the light ray escapes refractively
according to Snell's and Fresnel's laws. 

% Spectral statistics
An important topic of quantum chaos is the analysis of the statistical
properties of energy levels in quantum systems whose classical dynamics is
fully or partially chaotic~\cite{Stoeckmann00}. In the last years, the focus
has shifted from 
closed to open systems. Quantum eigenenergies of open systems 
(resonance frequencies in the case of microcavities) are complex valued with
the imaginary part being related to the lifetime of the state. 
Of particular interest is the fractal Weyl law for open chaotic
systems~\cite{Sjoestrand90,Zworski99,Lin02,LZ02,LSZ03,ST04,Nonnenmacher06}.
This is an extension of the well-known Weyl's formula for bounded systems 
which states that the number of levels with wave number $k_n \leq k$ is
asymptotically $N(k) \sim 
k^2$ for the particular case of a two-dimensional system which scales with the
energy such as quantum billiards. The fractal Weyl law for an open chaotic
system (which again scales with the energy) having complex wave 
numbers $k_n$ can be written as 
\begin{equation}\label{eq:Weyl}
N(k) = \{k_n: \imag{k_n} > -C,\; \real{k_n} \leq k\} \sim k^\alpha \ . 
\end{equation}
The cutoff constant $C > 0$ removes fast decaying states. It is
conjectured that the non-integer exponent is 
\begin{equation}\label{eq:alpha}
\alpha = \frac{D+1}{2} = \frac{d+2}{2}\ ,
 \end{equation}
where $D$ is the fractal dimension of the chaotic repeller of the open
chaotic system~\cite{LZ02}; $d=D-1$ is the dimension of the repeller in a
properly chosen surface of section~\cite{LichLieb92}.
The chaotic repeller is the
set of points in phase space that never leave the system both in forward and
in backward time dynamics, see, e.g., Ref.~\cite{LichLieb92}.

An essential prerequisite for a statistical analysis is a sufficient 
amount of resonance data. 
Resonance spectra can either be measured experimentally or computed by
exact quantum or semiclassical methods.
For some systems, e.g., atoms in external electric and magnetic fields
the complex $S$-matrix poles can be obtained directly by diagonalization
of non-Hermitian matrices using complex dilated basis sets \cite{Mai94,Tan96}.
However, in many theoretical calculations of, e.g., open billiards
\cite{Cvi89,Wir99,Mai02,Mai04}, optical microcavities \cite{Wiersig02b}, or
in the cycle-expanded Gutzwiller-Voros $\zeta$-function \cite{Vor88}
the resonances cannot be obtained directly by matrix diagonalization 
but rather by a numerically expensive root search
\begin{equation}\label{eq:det}
 \det{\bf M}(E_n) = 0 \; ,
\end{equation}
where the matrix ${\bf M}(E)$ depends on the complex valued energy $E$
(or on the complex wave number $k$ for billiards and microcavities).
The calculation of a real valued spectrum $\sigma(E)$ as the superposition
of Lorentzian line shapes is much less expensive, and it would be a great
advantage to be able to extract the complex $S$-matrix poles from a real
spectrum $\sigma(E)$.
In principle, this can be achieved by fitting the real valued spectrum to
a sum of Lorentzians.
Unfortunately, the established methods for fitting Lorentzians are 
restricted to isolated or weakly overlapping resonances, whereas
typical open systems in physically interesting regions are characterized
by strongly overlapping resonances.

In this paper we will apply the harmonic inversion method to obtain
the resonances of open dielectric microcavities.
We will show that the conjectured fractal Weyl law for open chaotic systems
is valid for microcavities only if the concept of the chaotic repeller is 
extended to a multifractal by incorporating Fresnel's laws.
The paper is organized as follows.
In Sec.~\ref{hi:sec} we present the numerical technique for the extraction
of the resonance poles from a real spectrum $\sigma(E)$.
In Sec.~\ref{ms:sec} we introduce the dielectric microstadium.
The high-index and low-index microstadium are investigated in detail
in Secs.~\ref{high:sec} and \ref{low:sec}.
Conclusion are given in Sec.~\ref{conclusion:sec}.

\section{The harmonic inversion technique}
\label{hi:sec}
Here, we will apply methods for high-resolution signal processing to
extract the resonance positions, widths, and amplitudes from a real spectrum
$\sigma(E)$.
In a first step the spectrum is Fourier transformed with appropriate 
frequency filters to obtain a band-limited time signal.
This time signal is analyzed, in the second step, by a high-resolution 
harmonic inversion method.
The harmonic inversion method for signal processing has been introduced
by Wall and Neuhauser \cite{Wal95} and then refined and improved by
Mandelshtam and Taylor \cite{Man97a,Man97b}.
It has proved to be a powerful tool for both the high-resolution analysis
of quantum spectra \cite{Mai97a} and the semiclassical quantization of
non-integrable systems \cite{Mai97b,Mai98,Mai99a,Mai99b}.
For a review on the use of the harmonic inversion method in semiclassical
physics see Ref.~\cite{Mai99c}.
Technically, the harmonic inversion problem can be recast as a set 
of nonlinear equations, which can be solved by either linear predictor,
Pad\'e approximant, or direct signal diagonalization \cite{Bel00a}.
Although details of the procedure have already been published we provide
here, for the convenience of the reader and to make the paper self-contained,
a brief description of the harmonic inversion method extended to the
extraction of resonances from experimental or theoretical spectra given as
superpositions of Lorentzians, i.e., spectra of the form
\begin{equation}
 f(\omega) = \sum_k \frac{A_k}{(\omega-\Omega_k)^2+(\Gamma_k/2)^2} \; ,
\label{f:eq}
\end{equation}
where the $\{\Omega_k,\Gamma_k,A_k\}$ are the frequency positions, widths,
and amplitudes of the resonances.
The frequency spectrum (\ref{f:eq}) can be interpreted as the Fourier
transform of a time signal
\begin{equation}
 C(t) = \sum_k d_k {\rm e}^{-{\rm i}\omega_k t} \; ,
\label{Ct:eq}
\end{equation}
with complex frequencies $\omega_k=\Omega_k-\frac{\rm i}{2}\Gamma_k$
and complex values $d_k$ simply related to the amplitudes $A_k$.
The challenge is to determine very broad resonances deep in the complex
plane and to resolve individual resonance poles in regions with strongly
overlapping resonances.
Basically, the harmonic inversion algorithm is split into the following
two steps.

In a first step, a frequency window 
$[\omega_0-\Delta\omega/2,\omega_0+\Delta\omega/2]$ is chosen.
A band-limited signal $C_{\rm bl}(t)$ with a finite number
of about 50 to 200 frequencies is obtained by the application of a 
windowed discrete Fourier transform to the frequency spectrum (\ref{f:eq}).
The band-limited signal is evaluated on an equidistant grid with time
step $\tau=2\pi/\Delta\omega$ and reads
\begin{equation}
   c_n \equiv C_{\rm bl}(t=n\tau)
 = \sum_{j=j_1}^{j_2} f(\omega_j) {\rm e}^{{\rm i} (\omega_j-\omega_0) n\tau} \; ,
\label{C_bl:eq}
\end{equation}
with $n =0,1,\dots,N_{\rm bl}-1$.
The limits $j_1$ and $j_2$ of the sum in Eq.~(\ref{C_bl:eq}) are taken
so that the equidistant grid points $\omega_j$ of the digitized spectrum
(\ref{f:eq}) cover the selected frequency filter
$[\omega_0-\Delta\omega/2,\omega_0+\Delta\omega/2]$.
The frequencies in the exponent of Eq.~(\ref{C_bl:eq}) are shifted by
$-\omega_0$ to reduce the phase oscillations of the band-limited signal.
The number $N_{\rm bl}$ of data points $c_n$ of the band-limited signal is
restricted by the number of grid points in the frequency window, i.e.,
$N_{\rm bl}\le j_2-j_1+1$.
We choose $N_{\rm bl}=2K$, where $K$ is an upper bound of the number of
resonances in the window (see below).

In a second step, the (shifted) frequencies $\omega_k'=\omega_k-\omega_0$
of the band-limited signal (\ref{C_bl:eq}) are obtained by solving the
nonlinear set of equations
\begin{equation}
 c_n = \sum_{k=1}^K d_k z_k^n \quad ; \quad n=0,1,\dots,2K-1
\label{cn_bl:eq}
\end{equation}
where $z_k\equiv\exp(-{\rm i}\omega_k'\tau)$ and $d_k$
are in general complex variational parameters.
As the number of frequencies in the band-limited signal is relatively small
($K\sim 50-200$), several variants of Prony's method \cite{Pro1795}, which
otherwise would be numerically unstable, can now be applied successfully,
e.g., linear predictor, Pad\'e approximant, or direct signal diagonalization 
\cite{Bel00a}.
Here, we briefly elaborate on the Pad\'e approximant.

Let us assume for the moment that the signal points $c_n$ are known up
to infinity, $n=0,1,\dots\infty$.
Interpreting the $c_n$'s as the coefficients of a Maclaurin series in the
variable $z^{-1}$, we can define the function
$g(z)=\sum_{n=0}^\infty c_n z^{-n}$.
With Eq.~(\ref{cn_bl:eq}) and the sum rule for geometric series
we obtain
\begin{eqnarray}
     g(z) &\equiv& \sum_{n=0}^\infty c_n z^{-n}
  =  \sum_{k=1}^K d_k \sum_{n=0}^\infty (z_k/z)^n \nonumber \\
 &=& \sum_{k=1}^K \frac{z d_k}{z-z_k} \equiv \frac{P_{K}(z)}{Q_K(z)} \; .
\label{g_Pade:eq}
\end{eqnarray}
The right-hand side of Eq.~(\ref{g_Pade:eq}) is a rational function 
with polynomials of degree $K$ in the numerator and denominator.
Evidently, the parameters $z_k=\exp{(-i\omega_k'\tau)}$ are the
poles of $g(z)$, i.e., the zeros of the polynomial $Q_K(z)$.
The parameters $d_k$ are calculated via the residues of the last two terms 
of Eq.~(\ref{g_Pade:eq}).
Application of the residue calculus yields
\begin{equation}
 d_k = \frac{P_{K}(z_k)}{z_k Q_K'(z_k)} \; ,
\label{dk_pade:eq}
\end{equation}
with the prime indicating the derivative ${\rm d}/{{\rm d}z}$.
Of course, the assumption that the coefficients $c_n$ are known up to
infinity is not fulfilled  and, therefore, the sum over all $c_nz^{-n}$
in Eq.~(\ref{g_Pade:eq}) cannot be evaluated in practice.
However, the convergence of the sum can be accelerated by use of
the Pad\'e approximant.
Indeed, knowledge of $2K$ signal points $c_0,\dots,c_{2K-1}$ is 
sufficient for the calculation of the coefficients of the two polynomials
\begin{equation}
 P_{K}(z) = \sum_{k=1}^{K} b_k z^k  \mbox{~~and~~}
 Q_K(z) = \sum_{k=1}^K a_k z^k - 1 \; .
\label{P_Q_polynomials:eq}
\end{equation}
The coefficients $a_k$ with $k=1,\dots,K$ are obtained as solutions of the 
linear set of equations
\begin{equation}
 c_n = \sum_{k=1}^K a_k c_{n+k} \; ,  \quad n = 0, \dots, K-1 \; .
\end{equation}
Once the $a_k$ are known, the coefficients $b_k$ are given by the explicit
formula
\begin{equation}
 b_k = \sum_{m=0}^{K-k} a_{k+m} c_{m} \; , \quad k = 1, \dots , K \; .
\end{equation}
Note that the Pad\'e approximant does not only accelerate the convergence
of the sum over all $c_nz^{-n}$ in Eq.~(\ref{g_Pade:eq}) but also yields
an analytic continuation for $z$ values, where the sum is not absolutely
convergent.

The parameters $z_k=\exp(-{\rm i}\omega_k'\tau)$ and thus the frequencies
\begin{equation}
 \omega_k = \omega_0 + \omega_k' = \omega_0 + \frac{\rm i}{\tau}\ln{z_k}
\end{equation}
are obtained by searching for the zeros of the polynomial $Q_K(z)$ 
in Eq.~(\ref{P_Q_polynomials:eq}).
This is the only nonlinear step of the algorithm.
The roots of polynomials can be found, in principle, by application
of Laguerre's method \cite{Press88}.
However, it turns out that an alternative method, i.e., the 
diagonalization of the Hessenberg matrix 
\begin{equation}
 {\bf A} =
 \left(\begin{array}{ccccc}
  -\frac{a_{K-1}}{a_K} & -\frac{a_{K-2}}{a_K} & \cdots &
  -\frac{a_{1}}{a_K} & -\frac{a_{0}}{a_K} \\
  1 & 0 & \cdots & 0 & 0 \\
  0 & 1 & \cdots & 0 & 0 \\
  \vdots & & & & \vdots \\
  0 & 0 & \cdots & 1 & 0 
 \end{array} \right) \quad ,
\label{Hesse:eq}
\end{equation}
with $a_k$ the coefficients of the polynomial $Q_K(z)$ in
Eq.~(\ref{P_Q_polynomials:eq}) is a numerically more robust technique for
finding the roots of high degree ($K \gtrsim 60$) polynomials \cite{Press88}.

When the harmonic inversion method is applied to a continuous spectrum
of a scattering system, given as a superposition of Lorentzian line shapes,
the resulting $\omega_k$ are the complex poles of the $S$-matrix.

\section{The microstadium}
\label{ms:sec}
% mode equation
Below a certain cutoff frequency $\omega_{\mbox{\footnotesize cutoff}}$,
microdisk-like 
cavities can be regarded as two-dimensional systems. In this case Maxwell's 
equations reduce to a two-dimensional scalar mode 
equation~\cite{Jackson83eng}
\begin{equation}\label{eq:wave}
-\nabla^2\psi = \neff^2(x,y)\frac{\omega^2}{c^2}\psi \ ,
\end{equation}
with effective index of refraction $n$, frequency $\omega$ and the speed of
light in vacuum $c$. 
The mode equation~(\ref{eq:wave}) is valid for both transverse magnetic (TM)
and transverse electric (TE) polarization. We focus on TM
polarization with the electric field $\vec{E}(x,y,t) =
(0,0,\psi(x,y)e^{-i\omega t})$ 
perpendicular to the cavity plane.  
The wave function $\psi$ and its normal derivative are continuous across the
boundary of the cavity. At infinity, outgoing wave conditions are imposed. 
Note that we ignore in the following a slight frequency dependence of $n$.

There are only a few reports in the literature on the spectral statistics of
microcavities. In a recent work the resonance width distribution in the
circular disk has been studied~\cite{RLRPK07}.
The integrability of this 
system allows to easily gather sufficient resonance data. In
Ref.~\cite{SJNS00} the resonance width distribution of a circular 
cavity with strong surface roughness has been analyzed. However, an additional
approximation was necessary in order to get a sufficient amount of resonances.

\begin{figure}[ht]
\includegraphics[width=0.65\figurewidth]{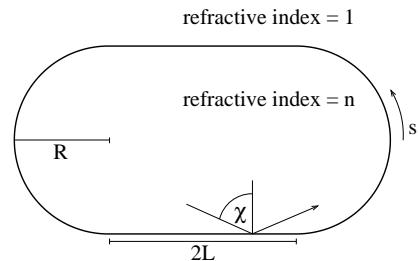}
\caption{Schematic top view of the microstadium cavity. Total internal
reflection of a ray with angle of incidence $\chi$ is illustrated. The
arclength coordinate along the circumference of the cavity's boundary is
denoted by $s$. }
\label{fig:sketch}
\end{figure}

In our approach, no such approximation is needed. Moreover, it is not
restricted to a particular geometry. As example, we consider the dielectric
microstadium as illustrated in Fig.~\ref{fig:sketch}. The cross-sectional area
is bounded by two semicircles with radius $R > 0$ and two straight lines of
length $2L > 0$. The corresponding stadium billiard is a paradigm for
fully chaotic systems~\cite{Bunimovich74b,Bunimovich79}. We choose the case
$L = R$ which is the most chaotic one~\cite{BenS78}. 

% previous work on spatial mode structure in stadium 
The spatial structure of optical modes in microstadiums has been extensively
studied in the context of ray-wave correspondence in open
systems~\cite{SHTS06,FHW06,LLZSB07,SH07,SFH07}. Our aim is to study the
spectral properties of modes in such a kind of cavity.

\section{High-index microstadium}
\label{high:sec}
% GaAs and polymer cavities
We consider first a GaAs microstadium with a high refractive index
$n=3.3$, the experimental realization of which has been reported in
Refs.~\cite{HF04,FYS07}. 

\subsection{Spectral analysis of resonances}
\begin{figure}[ht]
\includegraphics[width=\figurewidth]{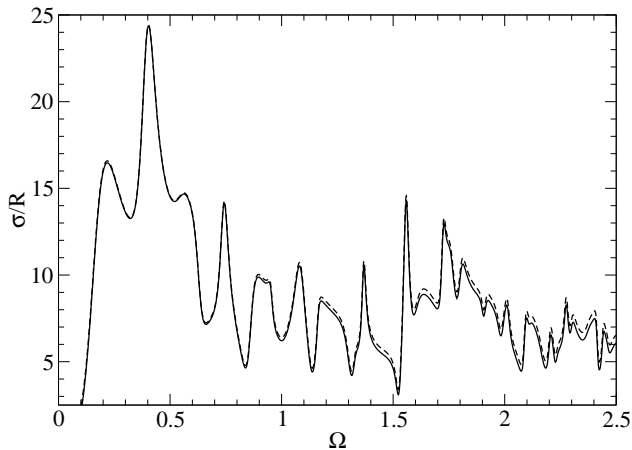}
\caption{Calculated total scattering cross section $\sigma$ versus real-valued
  frequency $\Omega = \omega R/c$ for the high-index microstadium.
The solid curve shows the result of 
the BEM with a plane wave incidence at $30^\circ$ to the horizontal
line segments of the stadium. The dashed curve is the reconstructed $\sigma$ 
based on the resonances extracted by the harmonic inversion.}  
\label{fig:sigma}
\end{figure}

% numerical scheme to find resonances
The modes~(\ref{eq:wave}) can be numerically computed with the boundary
element method (BEM) using a root search algorithm in the complex frequency 
plane~\cite{Wiersig02b}. Even though the BEM is very efficient, it does not 
allow for obtaining a sufficient amount of resonances for a statistical
analysis. Our improved strategy is therefore based on the BEM and the 
harmonic inversion technique. It consists of two steps. In a first step, we
compute spectral data on 
the real frequency axis. For convenience, we choose the total scattering cross
section $\sigma$ as function of the normalized frequency $\Omega = \omega R/c
= kR$ using the BEM. For a definition of the scattering cross
section we refer to Ref.~\cite{Wiersig02b}. We are able to cover the interval
$\Omega \in [0,25]$ where we found $3772$ resonances.
The upper bound $\Omega = 25$ corresponds to 
$R = 3.2\,\mu$m assuming a typical wavelength of $800\,$nm in
GaAs devices~\cite{HF04,FYS07}. To keep 
the picture clear, Fig.~\ref{fig:sigma} shows the calculated scattering cross
section in a smaller frequency interval. 

\begin{figure}[ht]
\includegraphics[width=\figurewidth]{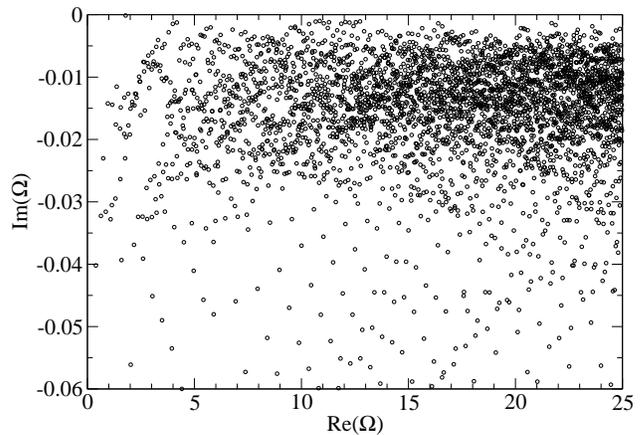}
\caption{Resonances in the plane of complex frequencies $\Omega$
  for the microstadium with $n=3.3$.
  We found $3772$ resonances in total.}
\label{fig:complexplaneA}
\end{figure}

In a second step we analyze the spectral data on the real frequency axis with
the help of the harmonic inversion technique.
Fig.~\ref{fig:complexplaneA} shows the resulting 
resonances in the complex plane with $\imag{\Omega} \geq -0.06$. The imaginary
part is related to the quality factor via $Q =-\real{\Omega}/[2\imag{\Omega}]$.
The limitations of the harmonic inversion method are as follows:
Very weak and broad resonances, i.e., resonances with small amplitudes
$A_k$ and large widths $\Gamma_k$ in Eq.~(\ref{f:eq}) may be overlooked.
Also, in rare cases one true resonance can be fitted by a sum of two
Lorentzians with nearly identical resonance positions and widths.
We have two ways to verify the harmonic inversion technique.
First, we can check exemplarily resonances with the BEM
by using root searching in the complex plane.
We always find good agreement.
Second, from the extracted resonances we can reconstruct the scattering
cross section. Fig.~\ref{fig:sigma} reveals good agreement with the
original scattering cross section. 

\begin{figure}[ht]
\includegraphics[width=\figurewidth]{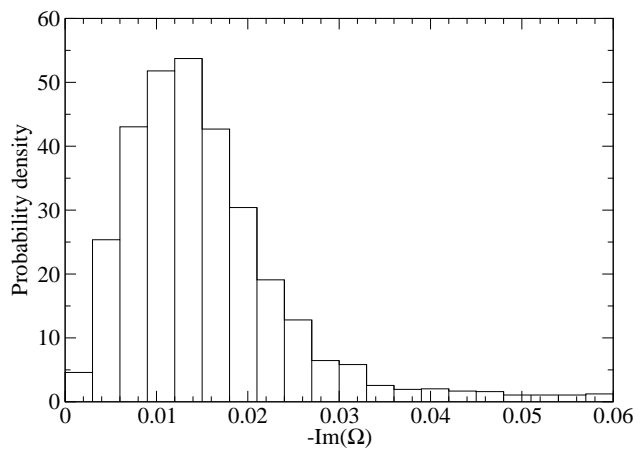}
\caption{Probability density of imaginary part of resonant frequencies $\Omega$ for the
  high-index microstadium.}
\label{fig:lifetimeA}
\end{figure}

The statistics of imaginary parts $\imag{\Omega} = -R/(2c\tau)$, where
$\tau$ is the lifetime of the given mode, is depicted in
\fig~\ref{fig:lifetimeA}. We see a clear maximum around $\imag{\Omega}
\approx -0.014$.

\begin{figure}[ht]
\includegraphics[width=\figurewidth]{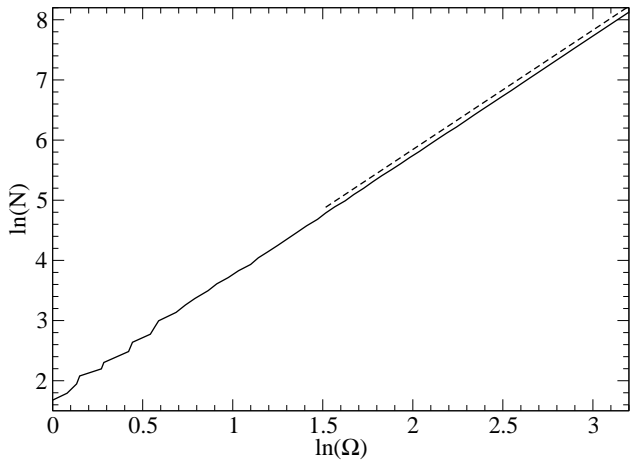}
\caption{Number of modes $N$ with cutoff $C = 0.06$ versus frequency $\Omega$
  in a double-logarithmic plot (solid line) for the microstadium with
  $n=3.3$. Dashed line with slope $\alpha \approx 1.98$ is a least-square fit;
  offset for clarity.}  
\label{fig:fitA}
\end{figure}

Figure~\ref{fig:fitA} shows the number of modes with cutoff $C = 0.06$
according to Eq.~(\ref{eq:Weyl}) versus normalized frequency $\Omega = kR$ in
a log-log 
plot. The data can be extremely well fitted by a straight line with slope
$\alpha \approx 1.98$. Because of the well pronounced maximum in the 
statistics of imaginary parts in Fig.~\ref{fig:lifetimeA} we expect only
a small change of the 
exponent $\alpha$  when the cutoff $C$ is varied. Indeed, this expectation is
confirmed by the numerics as we find $\alpha\in [1.96,2.02]$ for
$C\in[0.03,0.1]$. 
%Surprisingly, we recover, within the numerical accuracy, the exponent of the Weyl law for closed systems.
As an exponent larger than 2 for two-dimensional open systems is rather
unphysical, we believe that the regime $[2,2.02]$ is solely due to the
numerical uncertainty. 

\subsection{The chaotic repeller}
The chaotic repeller is the set of points in phase space that never leave the
system both in forward and in backward time dynamics~\cite{LichLieb92}. The
stable (unstable) manifold of a chaotic repeller 
is the set of points that converges to the repeller in forward (backward) time
evolution. The unstable manifold therefore describes the route of escape from
the chaotic system. For the case of chaotic microcavities it has been
demonstrated that the unstable manifold can play an important role for the 
far-field emission pattern~\cite{SRTCS04,LRKRK05,SH07,LYMLASLK07,WH07}.   

Here, we discuss the chaotic repeller of the microstadium restricted
to the Poincar\'e surface of section (SOS)~\cite{LichLieb92}. It is a plot of
the 
intersection points of a set of ray trajectories with a surface in phase
space. For billiards and microcavities it is convenient to take the cavity
boundary as SOS. The dimension of the repeller on this reduced phase
space is $d = D-1$. The reduced phase
space is parametrized by the arclength coordinate along
the circumference of the cavity's boundary, $s$, and the tangential momentum,
$p = \sin{\chi}$, where $\chi$ is the angle of incidence measured from the
surface normal; see Fig.~\ref{fig:sketch}. 

\begin{figure}[ht]
\includegraphics[width=0.9\figurewidth]{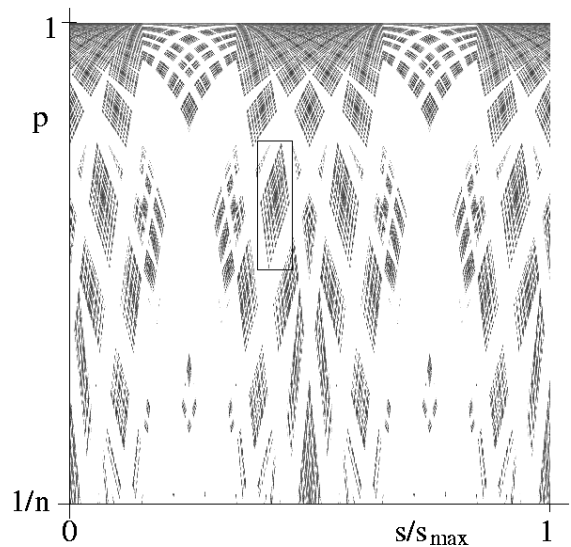}
\caption{Chaotic repeller (dark regions) of the high-index microstadium in the
  Poincar\'e surface of section. The horizontal
axis is the arclength coordinate $s/s_{\mbox{\footnotesize max}} \in [0,1]$ 
and the vertical axis is the tangential momentum $p = \sin{\chi} \in
[1/n,1]$. The rectangular region is magnified in Fig.~\ref{fig:repellerAmag}(a).}
\label{fig:repellerA}
\end{figure}

\begin{figure}[ht]
\includegraphics[width=0.9\figurewidth]{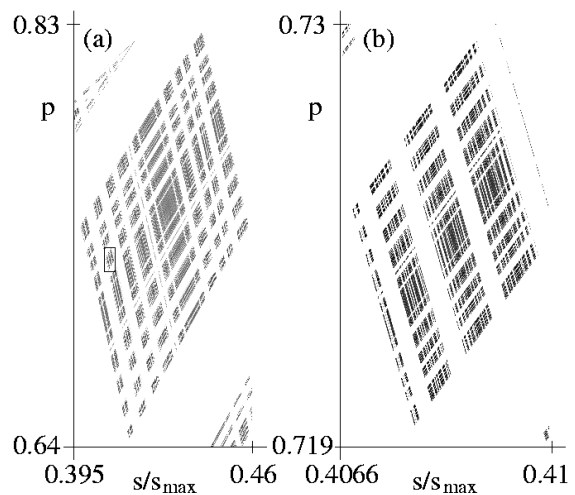}
\caption{(a) Magnification of rectangular region in
  Fig.~\ref{fig:repellerA}. (b)  Magnification of rectangular region in
  Fig.~\ref{fig:repellerAmag}(a). The fractal structure of the chaotic
  repeller can be clearly seen.}
\label{fig:repellerAmag}
\end{figure}

To compute the repeller we portion the phase space above the upper critical
line for total internal reflection, $p > |\sin{\chi_c}| = 1/n$, into
$L\times L = 1024\times 1024$ boxes. For each box we would like to determine whether it
belongs to the repeller or not. Obviously, this depends on the chosen
point within the box. To soften this arbitrariness the following averaging 
procedure is used. In each box $m = 400$ trajectories are started and
propagated both forward
and backward in time. Given a cutoff length $L_{\mbox{\footnotesize cutoff}}
= 20.5R$, we assign an intensity $I = 1$ to rays which never enter the leaky
region $|p| < 1/n$, otherwise $I = 0$. The weight $w\in [0,1]$ associated
with each box is then 
\begin{equation}\label{eq:weights}
w = \frac{1}{2m}\sum_{i=1}^{2m}I_i \ .
\end{equation}
Figure~\ref{fig:repellerA} shows the chaotic repeller determined in this
way. The magnifications in Fig.~\ref{fig:repellerAmag} demonstrate the fractal
nature of the repeller.

In order to compute the fractal dimension of the chaotic repeller, we apply the
thermodynamic formalism for fractal measures; see, e.g., Ref.~\cite{BS93}. 
As for the determination of the chaotic repeller we divide the region $(x,y) =
[s/s_{\mbox{\footnotesize max}}, (p-1/n)/(1-1/n)] 
\in[0,1]\times[0,1]$ into $L\times L = 1024\times 1024$
boxes with size $\varepsilon \times \varepsilon$ where $\varepsilon =
1/L$. The weights $w_{kl} = w(x_k,y_l)$ from Eq.~(\ref{eq:weights}) are normalized such
that $\sum_{kl} 
w_{kl} = 1$. We write the coarse-grained partition function as 
\begin{equation}\label{eq:Gamma}
\Gamma(\varepsilon,q) = \sum_{kl} w_{kl}^q \ ,
\end{equation}
where $q$ is a real-valued parameter. The scaling behaviour of the partition
function 
\begin{equation}\label{eq:scaling}
\Gamma(\varepsilon,q) \sim \varepsilon^{(q-1)d(q)}
\end{equation}
defines the generalized or R\'enyi dimensions $d(q)$. The dimension~$d(q)$ can
be determined numerically for given $q\neq 1$ by computing the partition
function $\Gamma$ as in Eq.~(\ref{eq:Gamma}) for increasing box sizes
$\varepsilon_M = 2^{-M}$ with $M = 10,9,8,7,6,5$. The weight of each resulting
box is then the sum of the weights of the corresponding smaller
boxes. Finally, the dimension $d$ is 
extracted by fitting the function~(\ref{eq:scaling}). The dimension $d(0)$ is 
equal to the Minkowski dimension, also called capacity or box counting
dimension. 
In the limit $q\to 1$, the R\'enyi dimension becomes the information dimension
\begin{equation}
d(1) = \lim_{\varepsilon\to 0} \frac{1}{\ln{\varepsilon}} \sum_{kl}
w_{kl}\ln{w_{kl}} \ .
\end{equation}
Similar to the general case of $d(q)$ with $q\neq 1$, $d(1)$ can be determined
by fitting $d(1)\ln{\varepsilon_M}$ to $\sum_{kl} w_{kl}\ln{w_{kl}}$ as function
of the box size $\varepsilon_M = 2^{-M}$.
The dimension $d(2)$ is called correlation dimension since it is related to
correlation functions as, e.g., in Ref.~\cite{Wiersig00}.

In the case of a fractal set with uniform measure, i.e., a point belongs to the
set ($I = 1$) or not ($I = 0$), all dimensions $d(q)$ have the same
value. In the more general case of fractal sets with nonuniform measure
($I\in\,\R^+$) the dimensions $d(q)$ differ. In 
such a case one speaks about a {\it multifractal}. The chaotic repeller is
strictly speaking a set with uniform measure. For numerical
reasons, however, we have seen that it is more convenient to consider a
real-valued $I(s,p)$. That this approximation is justified can be seen by the
fact that the numerically computed fractal dimensions $d(0)$, $d(1)$, and    
$d(2)$ are all approximately $1.68$. The predicted exponent according to
Eq.~(\ref{eq:alpha}) and $d = 1.68$ is then around $1.84$. 
Unexpectedly, this is not close to the Weyl exponent $\alpha\in [1.96,2.02]$
obtained from the counting of resonances. Hence, it seems that the conjectured
fractal Weyl law fails for the dielectric microstadium. 

% escape rate
In order to see the origin of this failure, let us consider the escape rate
$\gamma$ of the chaotic repeller. Points in the vicinity of the 
repeller escape exponentially in time with rate $\gamma$. To determine 
this rate, we remark that the numerically computed repeller in
Fig.~\ref{fig:repellerA} is an approximation to the real repeller, in that
numerically computed phase space points are never exactly located on the
repeller. We can exploit this fact to determine $\gamma$ by evolving the
numerically computed repeller in time. Its spatial structure does not change
under time evolution, only the total intensity decays exponentially. The
corresponding decay rate is nothing else than the escape rate $\gamma$.
% result
We find $\gamma \approx 0.056c/R$. This translates into $\imag{\Omega} =
-\gamma R/(2c) \approx -0.028$ which is significantly larger than the mean
escape rate of the quasi-bound modes in \fig~\ref{fig:lifetimeA}. We conclude
that the chaotic repeller as defined above also fails to describe the mean
lifetimes of the quasi-bound modes. 

In the following we show that both problems stem from the fact that the 
conventional chaotic repeller ignores the {\it partial escape} at dielectric
interfaces according to Fresnel's laws. Rays entering the leaky region are not
completely transmitted to the exterior of the cavity. For TM polarization the
reflection coefficient is  
\begin{equation}\label{eq:R}
R_{\text{TM}} = \left(\frac{\sin{(\chi-\chi_t)}}{\sin{(\chi+\chi_t)}}\right)^2
\end{equation}
with the angle of incidence $\chi$ and the angle of transmission $\chi_t$ 
related by Snell's law $n\sin\chi = \sin\chi_t$.
We can include Fresnel's laws in the concept of the chaotic repeller by
allowing for a real-valued intensity $I$ assigned to each ray. A similar
approach has been used to describe far-field emission pattern based on
unstable manifolds~\cite{SH07,WH07}.
Initially, we
choose $I = 1$ for a given ray, and whenever the ray enters the leaky region at
a phase space point $(s,p)$ we multiply the intensity~$I$ by the reflection
coefficient $R_{\text{TM}}(s,p)$. Finally, the ray's intensity
is $0\leq I \leq 1$. The chaotic repeller defined in this way is a set with
nonuniform measure $I(s,p)$; it is a multifractal. 

\begin{figure}[ht]
\includegraphics[width=0.9\figurewidth]{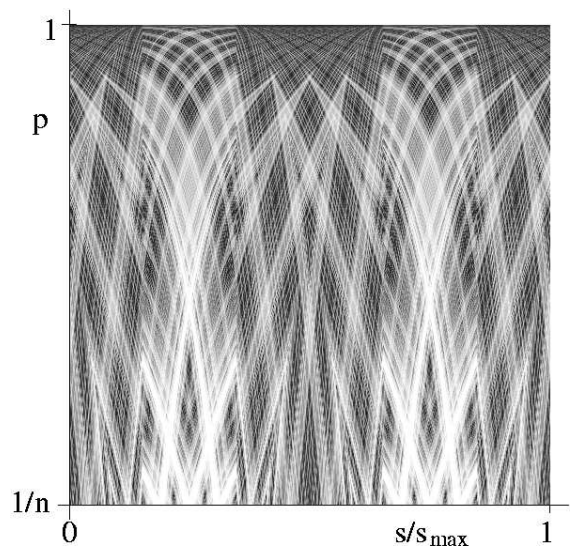}
\caption{Chaotic repeller of the microstadium with $n=3.3$ including Fresnel's
  laws; cf. Fig.~\ref{fig:repellerA}. Intensity is higher for darker regions,
  and vanishes in the white regions.}
\label{fig:repellerAfresnel}
\end{figure}

Using the extended concept of the chaotic repeller we compute
the escape rate to be $\gamma \approx 0.034c/R$. This gives
$\imag{\Omega} \approx -0.017$ which is in much better agreement with the
distribution of imaginary parts of the quasi-bound modes depicted in
\fig~\ref{fig:lifetimeA}. Figure~\ref{fig:repellerAfresnel} shows the chaotic
repeller including Fresnel's laws. The fractal
dimensions turn out to be $d(0) \approx 1.986$, $d(1) \approx 1.913$, and 
$d(2) \approx 1.877$. The predicted exponent for the Weyl law is therefore
$1.99$, $1.96$, and $1.94$, correspondingly. The values resulting from the
box-counting dimension and the information dimension are
in the interval of the Weyl exponent $\alpha\in [1.96,2.02]$ calculated from
the resonances. For the high-index microstadium we therefore have confirmed
the conjectured fractal Weyl law using the extended concept of the chaotic
repeller. 

%\clearpage
\section{Low-index microstadium}
\label{low:sec}
\begin{figure}[ht]
\includegraphics[width=\figurewidth]{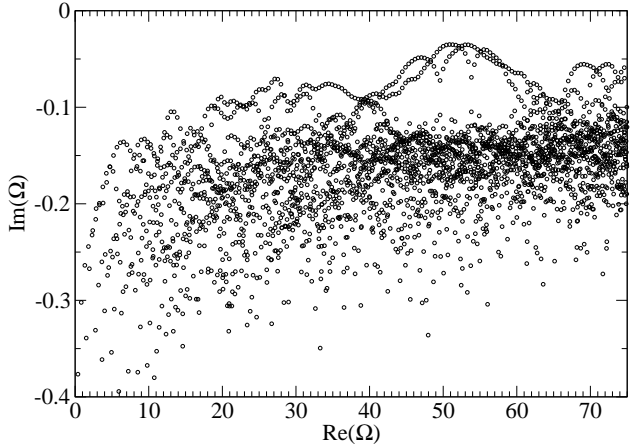}
\caption{Resonances in the complex-frequency plane for the microstadium with
  $n=1.5$. In total $2893$ resonances have been found.} 
\label{fig:complexplaneB}
\end{figure}

\begin{figure}[ht]
\includegraphics[width=\figurewidth]{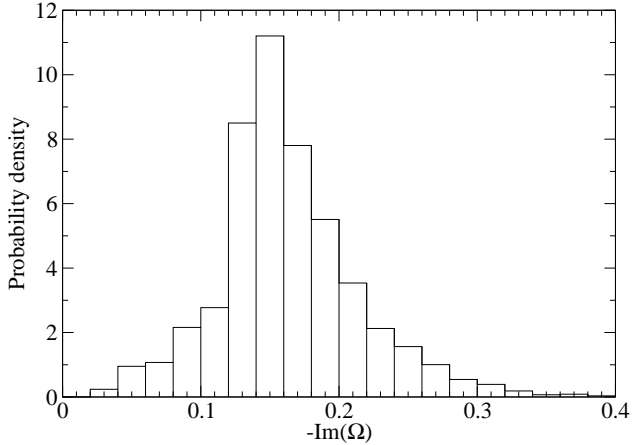}
\caption{Probability density of imaginary part of resonant frequencies $\Omega$ for the
  low-index microstadium.}
\label{fig:lifetimeB}
\end{figure}

Now we proceed to the microstadium with low refractive index $n=1.5$. 
Such a cavity has been fabricated recently by using a PMMA polymer
matrix~\cite{LLHZ06,LLZSB07}. 
Here, we can compute the resonances in the interval $\real{\Omega} \in [0,75]$
as can be observed in Fig.~\ref{fig:complexplaneB}. For example,
$\real{\Omega} = 75$ corresponds to $R = 7.2\;\mu$m for $\lambda =
600\;$nm~\cite{LLHZ06,LLZSB07}. The comparison with
Fig.~\ref{fig:complexplaneA} reveals that the resonances of the low-index
stadium are much deeper in the complex plane, which reflects the stronger
degree of openness due to the larger leaky region $|\sin{\chi_\text{c}}| =
1/n$. 
This can also be seen from the statistics of imaginary parts depicted in
\fig~\ref{fig:lifetimeB}. Here, we observe a clear maximum around 
$\imag{\Omega} \approx -0.15$.

\begin{figure}[t]
\includegraphics[width=\figurewidth]{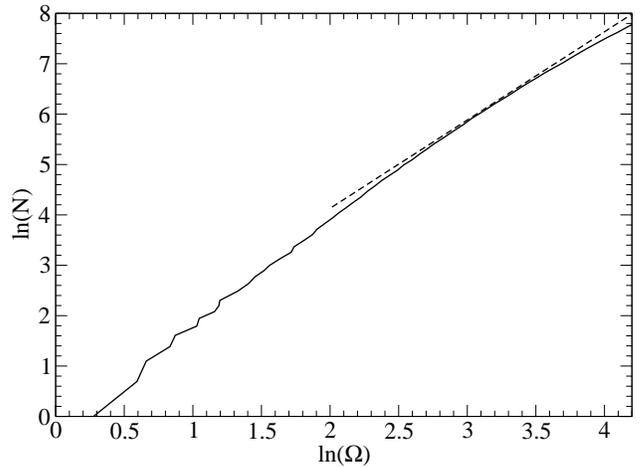}
\caption{Number of modes $N$ with cutoff $C = 0.3$ versus frequency $\Omega$
  in a double-logarithmic plot (solid line) for the microstadium with
  $n=1.5$. The slope of the linear fit is $\alpha \approx 1.75$ (dashed
  line; offset for clarity).}  
\label{fig:fitB}
\end{figure}

Figure~\ref{fig:fitB} shows a fit to the Weyl law~(\ref{eq:Weyl}) with $C =
0.3$ yielding $\alpha \approx 1.75$. Varying the cutoff parameter $C$ in the
interval $[0.25,0.4]$ we find $\alpha \in [1.68,1.88]$. For the strongly open
microstadium with low index of refraction we see a clear deviation from the
conventional Weyl law for closed systems as one would expect. 

\begin{figure}[ht]
\includegraphics[width=0.9\figurewidth]{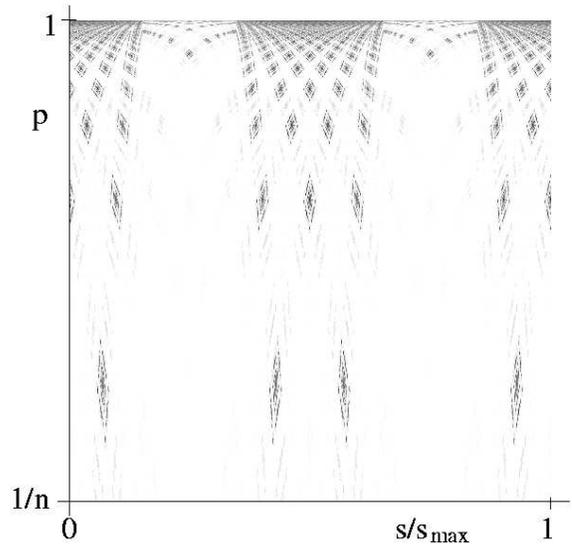}
\caption{Chaotic repeller of the low-index microstadium including Fresnel's
  laws.} 
\label{fig:repellerBfresnel}
\end{figure}

The chaotic repeller including Fresnel's laws is shown in
Fig.~\ref{fig:repellerBfresnel}. The escape rate for 
phase space points near the chaotic repeller is determined to be $\gamma
\approx 0.26c/R$. This translates into $\imag{\Omega} \approx -0.13$ in good
agreement with the distribution of the quasi-bound modes in
\fig~\ref{fig:lifetimeB}. 
Note that Fresnel's laws do not change much here since the reflectivity
in the leaky region is in general small due to the low refractive index.
% dimensions
The numerically computed dimensions are $d(0)\approx  1.512$ and $d(1)\approx
d(2) \approx 1.593$. Hence, the predicted exponent is $1.76$ and $1.78$ 
which is fully consistent with the fractal Weyl law with exponent $\alpha \in
[1.68,1.88]$.  

\section{Conclusion}
\label{conclusion:sec}
We have demonstrated that the harmonic inversion technique allows to compute
very efficiently a huge amount of complex resonances in dielectric
microcavities.  
This is of high practical value for the statistical analysis of resonances 
in the context of quantum chaos and complex scattering. The
approach has been illustrated for the chaotic microstadium made of GaAs and
polymers. We demonstrate that the fractal Weyl law can be applied to the class
of dielectric cavities by extending the conventional concept of the chaotic
repeller by including partial escape of light according to Fresnel's laws. 
We expect that our finding is of high relevance for the understanding of
ray-wave correspondence in chaotic microcavities. Moreover, it can have
implications on multi-mode lasing in such cavities~\cite{SHI05}. 

\section{Acknowledgement}
We would like to thank E.~Bogomolny, M.~Hentschel, and S.~Shinohara for
discussions. Financial support from the DFG research group 760 is acknowledged.

%\clearpage
%\bibliography{bib}
\bibliography{}

\begin{thebibliography}{59}
\expandafter\ifx\csname natexlab\endcsname\relax\def\natexlab#1{#1}\fi
\expandafter\ifx\csname bibnamefont\endcsname\relax
  \def\bibnamefont#1{#1}\fi
\expandafter\ifx\csname bibfnamefont\endcsname\relax
  \def\bibfnamefont#1{#1}\fi
\expandafter\ifx\csname citenamefont\endcsname\relax
  \def\citenamefont#1{#1}\fi
\expandafter\ifx\csname url\endcsname\relax
  \def\url#1{\texttt{#1}}\fi
\expandafter\ifx\csname urlprefix\endcsname\relax\def\urlprefix{URL }\fi
\providecommand{\bibinfo}[2]{#2}
\providecommand{\eprint}[2][]{\url{#2}}

\bibitem[{\citenamefont{Park et~al.}(2004)\citenamefont{Park, Kim, Kwon, Ju,
  Yang, Baek, Kim, and Lee}}]{Park04}
\bibinfo{author}{\bibfnamefont{H.-G.} \bibnamefont{Park}},
  \bibinfo{author}{\bibfnamefont{S.-H.} \bibnamefont{Kim}},
  \bibinfo{author}{\bibfnamefont{S.-H.} \bibnamefont{Kwon}},
  \bibinfo{author}{\bibfnamefont{Y.-G.} \bibnamefont{Ju}},
  \bibinfo{author}{\bibfnamefont{J.-K.} \bibnamefont{Yang}},
  \bibinfo{author}{\bibfnamefont{J.-H.} \bibnamefont{Baek}},
  \bibinfo{author}{\bibfnamefont{S.-B.} \bibnamefont{Kim}}, \bibnamefont{and}
  \bibinfo{author}{\bibfnamefont{Y.-H.} \bibnamefont{Lee}},
  \bibinfo{journal}{Science} \textbf{\bibinfo{volume}{305}},
  \bibinfo{pages}{1444} (\bibinfo{year}{2004}).

\bibitem[{\citenamefont{Ulrich et~al.}(2007)\citenamefont{Ulrich, Gies, Ates,
  Wiersig, Reitzenstein, Hofmann, \protect{L\"offler}, Forchel, Jahnke, and
  Michler}}]{Ulrich06}
\bibinfo{author}{\bibfnamefont{S.~M.} \bibnamefont{Ulrich}},
  \bibinfo{author}{\bibfnamefont{C.}~\bibnamefont{Gies}},
  \bibinfo{author}{\bibfnamefont{S.}~\bibnamefont{Ates}},
  \bibinfo{author}{\bibfnamefont{J.}~\bibnamefont{Wiersig}},
  \bibinfo{author}{\bibfnamefont{S.}~\bibnamefont{Reitzenstein}},
  \bibinfo{author}{\bibfnamefont{C.}~\bibnamefont{Hofmann}},
  \bibinfo{author}{\bibfnamefont{A.}~\bibnamefont{\protect{L\"offler}}},
  \bibinfo{author}{\bibfnamefont{A.}~\bibnamefont{Forchel}},
  \bibinfo{author}{\bibfnamefont{F.}~\bibnamefont{Jahnke}}, \bibnamefont{and}
  \bibinfo{author}{\bibfnamefont{P.}~\bibnamefont{Michler}},
  \bibinfo{journal}{Phys. Rev. Lett.} \textbf{\bibinfo{volume}{98}},
  \bibinfo{pages}{043906} (\bibinfo{year}{2007}).

\bibitem[{\citenamefont{Kim et~al.}(1999)\citenamefont{Kim, Benson, Kan, and
  Yamamoto}}]{KBKY99}
\bibinfo{author}{\bibfnamefont{J.}~\bibnamefont{Kim}},
  \bibinfo{author}{\bibfnamefont{O.}~\bibnamefont{Benson}},
  \bibinfo{author}{\bibfnamefont{H.}~\bibnamefont{Kan}}, \bibnamefont{and}
  \bibinfo{author}{\bibfnamefont{Y.}~\bibnamefont{Yamamoto}},
  \bibinfo{journal}{Nature} \textbf{\bibinfo{volume}{397}},
  \bibinfo{pages}{500} (\bibinfo{year}{1999}).

\bibitem[{\citenamefont{Michler et~al.}(2000)\citenamefont{Michler,
  Imamo\u{g}lu, Mason, Carson, Strouse, and Buratto}}]{Michler2000}
\bibinfo{author}{\bibfnamefont{P.}~\bibnamefont{Michler}},
  \bibinfo{author}{\bibfnamefont{A.}~\bibnamefont{Imamo\u{g}lu}},
  \bibinfo{author}{\bibfnamefont{M.~D.} \bibnamefont{Mason}},
  \bibinfo{author}{\bibfnamefont{P.~J.} \bibnamefont{Carson}},
  \bibinfo{author}{\bibfnamefont{G.~F.} \bibnamefont{Strouse}},
  \bibnamefont{and} \bibinfo{author}{\bibfnamefont{S.~K.}
  \bibnamefont{Buratto}}, \bibinfo{journal}{Nature (London)}
  \textbf{\bibinfo{volume}{406}}, \bibinfo{pages}{968} (\bibinfo{year}{2000}).

\bibitem[{\citenamefont{Benyoucef et~al.}(2005)\citenamefont{Benyoucef, Ulrich,
  Michler, Wiersig, Jahnke, and Forchel}}]{BUMWJF04b}
\bibinfo{author}{\bibfnamefont{M.}~\bibnamefont{Benyoucef}},
  \bibinfo{author}{\bibfnamefont{S.}~\bibnamefont{Ulrich}},
  \bibinfo{author}{\bibfnamefont{P.}~\bibnamefont{Michler}},
  \bibinfo{author}{\bibfnamefont{J.}~\bibnamefont{Wiersig}},
  \bibinfo{author}{\bibfnamefont{F.}~\bibnamefont{Jahnke}}, \bibnamefont{and}
  \bibinfo{author}{\bibfnamefont{A.}~\bibnamefont{Forchel}},
  \bibinfo{journal}{J. Appl. Phys.} \textbf{\bibinfo{volume}{97}},
  \bibinfo{pages}{023101} (\bibinfo{year}{2005}).

\bibitem[{\citenamefont{N{\"o}ckel and Stone}(1997)}]{ND97}
\bibinfo{author}{\bibfnamefont{J.~U.} \bibnamefont{N{\"o}ckel}}
  \bibnamefont{and} \bibinfo{author}{\bibfnamefont{A.~D.} \bibnamefont{Stone}},
  \bibinfo{journal}{Nature (London)} \textbf{\bibinfo{volume}{385}},
  \bibinfo{pages}{45} (\bibinfo{year}{1997}).

\bibitem[{\citenamefont{Gmachl et~al.}(1998)\citenamefont{Gmachl, Capasso,
  Narimanov, N{\"o}ckel, Stone, Faist, Sivco, and Cho}}]{GCNNSFSC98}
\bibinfo{author}{\bibfnamefont{C.}~\bibnamefont{Gmachl}},
  \bibinfo{author}{\bibfnamefont{F.}~\bibnamefont{Capasso}},
  \bibinfo{author}{\bibfnamefont{E.~E.} \bibnamefont{Narimanov}},
  \bibinfo{author}{\bibfnamefont{J.~U.} \bibnamefont{N{\"o}ckel}},
  \bibinfo{author}{\bibfnamefont{A.~D.} \bibnamefont{Stone}},
  \bibinfo{author}{\bibfnamefont{J.}~\bibnamefont{Faist}},
  \bibinfo{author}{\bibfnamefont{D.~L.} \bibnamefont{Sivco}}, \bibnamefont{and}
  \bibinfo{author}{\bibfnamefont{A.~Y.} \bibnamefont{Cho}},
  \bibinfo{journal}{Science} \textbf{\bibinfo{volume}{280}},
  \bibinfo{pages}{1556} (\bibinfo{year}{1998}).

\bibitem[{\citenamefont{Berry}(1981)}]{Berry81}
\bibinfo{author}{\bibfnamefont{M.~V.} \bibnamefont{Berry}},
  \bibinfo{journal}{Eur. J. Phys.} \textbf{\bibinfo{volume}{2}},
  \bibinfo{pages}{91} (\bibinfo{year}{1981}).

\bibitem[{\citenamefont{Robnik}(1983)}]{Robnik83}
\bibinfo{author}{\bibfnamefont{M.}~\bibnamefont{Robnik}}, \bibinfo{journal}{J.
  Phys. A} \textbf{\bibinfo{volume}{16}}, \bibinfo{pages}{3971}
  (\bibinfo{year}{1983}).

\bibitem[{\citenamefont{St{\"o}ckmann}(2000)}]{Stoeckmann00}
\bibinfo{author}{\bibfnamefont{H.-J.} \bibnamefont{St{\"o}ckmann}},
  \emph{\bibinfo{title}{Quantum Chaos}} (\bibinfo{publisher}{Cambridge
  University Press}, \bibinfo{address}{Cambridge}, \bibinfo{year}{2000}).

\bibitem[{\citenamefont{Sj{\"o}strand}(1990)}]{Sjoestrand90}
\bibinfo{author}{\bibfnamefont{J.}~\bibnamefont{Sj{\"o}strand}},
  \bibinfo{journal}{Duke Math.} \textbf{\bibinfo{volume}{60}},
  \bibinfo{pages}{1} (\bibinfo{year}{1990}).

\bibitem[{\citenamefont{Zworski}(1999)}]{Zworski99}
\bibinfo{author}{\bibfnamefont{M.}~\bibnamefont{Zworski}},
  \bibinfo{journal}{Inv. Math.} \textbf{\bibinfo{volume}{136}},
  \bibinfo{pages}{353} (\bibinfo{year}{1999}).

\bibitem[{\citenamefont{Lin}(2002)}]{Lin02}
\bibinfo{author}{\bibfnamefont{K.~K.} \bibnamefont{Lin}}, \bibinfo{journal}{J.
  Comput. Phys.} \textbf{\bibinfo{volume}{176}}, \bibinfo{pages}{295}
  (\bibinfo{year}{2002}).

\bibitem[{\citenamefont{Lin and Zworski}(2002)}]{LZ02}
\bibinfo{author}{\bibfnamefont{K.~K.} \bibnamefont{Lin}} \bibnamefont{and}
  \bibinfo{author}{\bibfnamefont{M.}~\bibnamefont{Zworski}},
  \bibinfo{journal}{Chem. Phys. Lett.} \textbf{\bibinfo{volume}{355}},
  \bibinfo{pages}{201} (\bibinfo{year}{2002}).

\bibitem[{\citenamefont{Lu et~al.}(2003)\citenamefont{Lu, Sridhar, and
  Zworski}}]{LSZ03}
\bibinfo{author}{\bibfnamefont{W.~T.} \bibnamefont{Lu}},
  \bibinfo{author}{\bibfnamefont{S.}~\bibnamefont{Sridhar}}, \bibnamefont{and}
  \bibinfo{author}{\bibfnamefont{M.}~\bibnamefont{Zworski}},
  \bibinfo{journal}{Phys. Rev. Lett.} \textbf{\bibinfo{volume}{91}},
  \bibinfo{pages}{154101} (\bibinfo{year}{2003}).

\bibitem[{\citenamefont{Schomerus and Tworzydlo}(2004)}]{ST04}
\bibinfo{author}{\bibfnamefont{H.}~\bibnamefont{Schomerus}} \bibnamefont{and}
  \bibinfo{author}{\bibfnamefont{J.}~\bibnamefont{Tworzydlo}},
  \bibinfo{journal}{Phys. Rev. Lett.} \textbf{\bibinfo{volume}{93}},
  \bibinfo{pages}{154102} (\bibinfo{year}{2004}).

\bibitem[{\citenamefont{Nonnenmacher}(2006)}]{Nonnenmacher06}
\bibinfo{author}{\bibfnamefont{S.}~\bibnamefont{Nonnenmacher}}, in
  \emph{\bibinfo{booktitle}{Mathematical Physics of Quantum Mechanics}}, edited
  by \bibinfo{editor}{\bibfnamefont{J.}~\bibnamefont{Asch}} \bibnamefont{and}
  \bibinfo{editor}{\bibfnamefont{A.}~\bibnamefont{Joye}}
  (\bibinfo{publisher}{Springer}, \bibinfo{address}{Berlin},
  \bibinfo{year}{2006}), vol. \bibinfo{volume}{No.\ 690} of
  \emph{\bibinfo{series}{Springer Lecture Notes in Physics}}, pp.
  \bibinfo{pages}{435--450}.

\bibitem[{\citenamefont{Lichtenberg and Lieberman}(1992)}]{LichLieb92}
\bibinfo{author}{\bibfnamefont{A.~J.} \bibnamefont{Lichtenberg}}
  \bibnamefont{and} \bibinfo{author}{\bibfnamefont{M.~A.}
  \bibnamefont{Lieberman}}, \emph{\bibinfo{title}{Regular and Chaotic
  Dynamics}} (\bibinfo{publisher}{Springer}, \bibinfo{address}{Berlin},
  \bibinfo{year}{1992}).

\bibitem[{\citenamefont{Main and Wunner}(1994)}]{Mai94}
\bibinfo{author}{\bibfnamefont{J.}~\bibnamefont{Main}} \bibnamefont{and}
  \bibinfo{author}{\bibfnamefont{G.}~\bibnamefont{Wunner}},
  \bibinfo{journal}{J. Phys. B} \textbf{\bibinfo{volume}{27}},
  \bibinfo{pages}{2835} (\bibinfo{year}{1994}).

\bibitem[{\citenamefont{Tanner et~al.}(1996)\citenamefont{Tanner, Hansen, and
  Main}}]{Tan96}
\bibinfo{author}{\bibfnamefont{G.}~\bibnamefont{Tanner}},
  \bibinfo{author}{\bibfnamefont{K.~T.} \bibnamefont{Hansen}},
  \bibnamefont{and} \bibinfo{author}{\bibfnamefont{J.}~\bibnamefont{Main}},
  \bibinfo{journal}{Nonlinearity} \textbf{\bibinfo{volume}{9}},
  \bibinfo{pages}{1641} (\bibinfo{year}{1996}).

\bibitem[{\citenamefont{Cvitanovi{\'c} and Eckhardt}(1989)}]{Cvi89}
\bibinfo{author}{\bibfnamefont{P.}~\bibnamefont{Cvitanovi{\'c}}}
  \bibnamefont{and} \bibinfo{author}{\bibfnamefont{B.}~\bibnamefont{Eckhardt}},
  \bibinfo{journal}{Phys. Rev. Lett.} \textbf{\bibinfo{volume}{63}},
  \bibinfo{pages}{823} (\bibinfo{year}{1989}).

\bibitem[{\citenamefont{Wirzba}(1999)}]{Wir99}
\bibinfo{author}{\bibfnamefont{A.}~\bibnamefont{Wirzba}}, \bibinfo{journal}{J.
  Phys.A} \textbf{\bibinfo{volume}{309}}, \bibinfo{pages}{1}
  (\bibinfo{year}{1999}).

\bibitem[{\citenamefont{Main et~al.}(2002)\citenamefont{Main, Wunner,
  \protect{At\i lgan}, Taylor, and Dando}}]{Mai02}
\bibinfo{author}{\bibfnamefont{J.}~\bibnamefont{Main}},
  \bibinfo{author}{\bibfnamefont{G.}~\bibnamefont{Wunner}},
  \bibinfo{author}{\bibfnamefont{E.}~\bibnamefont{\protect{At\i lgan}}},
  \bibinfo{author}{\bibfnamefont{H.~S.} \bibnamefont{Taylor}},
  \bibnamefont{and} \bibinfo{author}{\bibfnamefont{P.~A.} \bibnamefont{Dando}},
  \bibinfo{journal}{Phys. Lett. A} \textbf{\bibinfo{volume}{305}},
  \bibinfo{pages}{176} (\bibinfo{year}{2002}).

\bibitem[{\citenamefont{Main et~al.}(2004)\citenamefont{Main, \protect{At\i
  lgan}, Taylor, and Wunner}}]{Mai04}
\bibinfo{author}{\bibfnamefont{J.}~\bibnamefont{Main}},
  \bibinfo{author}{\bibfnamefont{E.}~\bibnamefont{\protect{At\i lgan}}},
  \bibinfo{author}{\bibfnamefont{H.~S.} \bibnamefont{Taylor}},
  \bibnamefont{and} \bibinfo{author}{\bibfnamefont{G.}~\bibnamefont{Wunner}},
  \bibinfo{journal}{Phys. Rev. E} \textbf{\bibinfo{volume}{69}},
  \bibinfo{pages}{056227} (\bibinfo{year}{2004}).

\bibitem[{\citenamefont{Wiersig}(2003)}]{Wiersig02b}
\bibinfo{author}{\bibfnamefont{J.}~\bibnamefont{Wiersig}}, \bibinfo{journal}{J.
  Opt. A: Pure Appl. Opt.} \textbf{\bibinfo{volume}{5}}, \bibinfo{pages}{53}
  (\bibinfo{year}{2003}).

\bibitem[{\citenamefont{Voros}(1988)}]{Vor88}
\bibinfo{author}{\bibfnamefont{A.}~\bibnamefont{Voros}}, \bibinfo{journal}{J.
  Phys. A} \textbf{\bibinfo{volume}{21}}, \bibinfo{pages}{685}
  (\bibinfo{year}{1988}).

\bibitem[{\citenamefont{Wall and Neuhauser}(1995)}]{Wal95}
\bibinfo{author}{\bibfnamefont{M.~R.} \bibnamefont{Wall}} \bibnamefont{and}
  \bibinfo{author}{\bibfnamefont{D.}~\bibnamefont{Neuhauser}},
  \bibinfo{journal}{J. Chem. Phys.} \textbf{\bibinfo{volume}{102}},
  \bibinfo{pages}{8011} (\bibinfo{year}{1995}).

\bibitem[{\citenamefont{Mandelshtam and Taylor}(1997{\natexlab{a}})}]{Man97a}
\bibinfo{author}{\bibfnamefont{V.~A.} \bibnamefont{Mandelshtam}}
  \bibnamefont{and} \bibinfo{author}{\bibfnamefont{H.~S.}
  \bibnamefont{Taylor}}, \bibinfo{journal}{Phys. Rev. Lett.}
  \textbf{\bibinfo{volume}{78}}, \bibinfo{pages}{3274}
  (\bibinfo{year}{1997}{\natexlab{a}}).

\bibitem[{\citenamefont{Mandelshtam and Taylor}(1997{\natexlab{b}})}]{Man97b}
\bibinfo{author}{\bibfnamefont{V.~A.} \bibnamefont{Mandelshtam}}
  \bibnamefont{and} \bibinfo{author}{\bibfnamefont{H.~S.}
  \bibnamefont{Taylor}}, \bibinfo{journal}{J. Chem. Phys.}
  \textbf{\bibinfo{volume}{107}}, \bibinfo{pages}{6756}
  (\bibinfo{year}{1997}{\natexlab{b}}).

\bibitem[{\citenamefont{Main et~al.}(1997{\natexlab{a}})\citenamefont{Main,
  Mandelshtam, and Taylor}}]{Mai97a}
\bibinfo{author}{\bibfnamefont{J.}~\bibnamefont{Main}},
  \bibinfo{author}{\bibfnamefont{V.~A.} \bibnamefont{Mandelshtam}},
  \bibnamefont{and} \bibinfo{author}{\bibfnamefont{H.~S.}
  \bibnamefont{Taylor}}, \bibinfo{journal}{Phys. Rev. Lett.}
  \textbf{\bibinfo{volume}{78}}, \bibinfo{pages}{4351}
  (\bibinfo{year}{1997}{\natexlab{a}}).

\bibitem[{\citenamefont{Main et~al.}(1997{\natexlab{b}})\citenamefont{Main,
  Mandelshtam, and Taylor}}]{Mai97b}
\bibinfo{author}{\bibfnamefont{J.}~\bibnamefont{Main}},
  \bibinfo{author}{\bibfnamefont{V.~A.} \bibnamefont{Mandelshtam}},
  \bibnamefont{and} \bibinfo{author}{\bibfnamefont{H.~S.}
  \bibnamefont{Taylor}}, \bibinfo{journal}{Phys. Rev. Lett.}
  \textbf{\bibinfo{volume}{79}}, \bibinfo{pages}{825}
  (\bibinfo{year}{1997}{\natexlab{b}}).

\bibitem[{\citenamefont{Main et~al.}(1998)\citenamefont{Main, Mandelshtam,
  Wunner, and Taylor}}]{Mai98}
\bibinfo{author}{\bibfnamefont{J.}~\bibnamefont{Main}},
  \bibinfo{author}{\bibfnamefont{V.~A.} \bibnamefont{Mandelshtam}},
  \bibinfo{author}{\bibfnamefont{G.}~\bibnamefont{Wunner}}, \bibnamefont{and}
  \bibinfo{author}{\bibfnamefont{H.~S.} \bibnamefont{Taylor}},
  \bibinfo{journal}{Nonlinearity} \textbf{\bibinfo{volume}{11}},
  \bibinfo{pages}{1015} (\bibinfo{year}{1998}).

\bibitem[{\citenamefont{Main and Wunner}(1999{\natexlab{a}})}]{Mai99a}
\bibinfo{author}{\bibfnamefont{J.}~\bibnamefont{Main}} \bibnamefont{and}
  \bibinfo{author}{\bibfnamefont{G.}~\bibnamefont{Wunner}},
  \bibinfo{journal}{Phys. Rev. Lett.} \textbf{\bibinfo{volume}{82}},
  \bibinfo{pages}{3038} (\bibinfo{year}{1999}{\natexlab{a}}).

\bibitem[{\citenamefont{Main and Wunner}(1999{\natexlab{b}})}]{Mai99b}
\bibinfo{author}{\bibfnamefont{J.}~\bibnamefont{Main}} \bibnamefont{and}
  \bibinfo{author}{\bibfnamefont{G.}~\bibnamefont{Wunner}},
  \bibinfo{journal}{Phys. Rev. A} \textbf{\bibinfo{volume}{59}},
  \bibinfo{pages}{R2548} (\bibinfo{year}{1999}{\natexlab{b}}).

\bibitem[{\citenamefont{Main}(1999)}]{Mai99c}
\bibinfo{author}{\bibfnamefont{J.}~\bibnamefont{Main}}, \bibinfo{journal}{Phys.
  Rep.} \textbf{\bibinfo{volume}{316}}, \bibinfo{pages}{233}
  (\bibinfo{year}{1999}).

\bibitem[{\citenamefont{\protect{D\v z}. Belki{\'c}
  et~al.}(2000)\citenamefont{\protect{D\v z}. Belki{\'c}, Dando, Main, and
  Taylor}}]{Bel00a}
\bibinfo{author}{\bibnamefont{\protect{D\v z}. Belki{\'c}}},
  \bibinfo{author}{\bibfnamefont{P.~A.} \bibnamefont{Dando}},
  \bibinfo{author}{\bibfnamefont{J.}~\bibnamefont{Main}}, \bibnamefont{and}
  \bibinfo{author}{\bibfnamefont{H.~S.} \bibnamefont{Taylor}},
  \bibinfo{journal}{J. Chem. Phys.} \textbf{\bibinfo{volume}{113}},
  \bibinfo{pages}{6542} (\bibinfo{year}{2000}).

\bibitem[{\citenamefont{de~Prony}(1795)}]{Pro1795}
\bibinfo{author}{\bibfnamefont{B.~G.~R.} \bibnamefont{de~Prony}},
  \bibinfo{journal}{Journal de l'\'Ecole Polytechnique}
  \textbf{\bibinfo{volume}{1}}, \bibinfo{pages}{24} (\bibinfo{year}{1795}).

\bibitem[{\citenamefont{Press et~al.}(1988)\citenamefont{Press, Flannery,
  Teukolsky, and Vetterling}}]{Press88}
\bibinfo{author}{\bibfnamefont{W.~H.} \bibnamefont{Press}},
  \bibinfo{author}{\bibfnamefont{B.~P.} \bibnamefont{Flannery}},
  \bibinfo{author}{\bibfnamefont{S.~A.} \bibnamefont{Teukolsky}},
  \bibnamefont{and} \bibinfo{author}{\bibfnamefont{W.~T.}
  \bibnamefont{Vetterling}}, \emph{\bibinfo{title}{Numerical Recipes in C. The
  Art of Scientific Computing.}} (\bibinfo{publisher}{Cambridge University
  Press}, \bibinfo{address}{Cambridge}, \bibinfo{year}{1988}).

\bibitem[{\citenamefont{Jackson}(1962)}]{Jackson83eng}
\bibinfo{author}{\bibfnamefont{J.~D.} \bibnamefont{Jackson}},
  \emph{\bibinfo{title}{Classical Electrodynamics}} (\bibinfo{publisher}{John
  Wiley and Sons Inc.}, \bibinfo{address}{New York, London},
  \bibinfo{year}{1962}).

\bibitem[{\citenamefont{Ryu et~al.}(2007)\citenamefont{Ryu, Lee, Rim, Park, and
  Kim}}]{RLRPK07}
\bibinfo{author}{\bibfnamefont{J.-W.} \bibnamefont{Ryu}},
  \bibinfo{author}{\bibfnamefont{S.-Y.} \bibnamefont{Lee}},
  \bibinfo{author}{\bibfnamefont{S.}~\bibnamefont{Rim}},
  \bibinfo{author}{\bibfnamefont{Y.-J.} \bibnamefont{Park}}, \bibnamefont{and}
  \bibinfo{author}{\bibfnamefont{C.-M.} \bibnamefont{Kim}}
  (\bibinfo{year}{2007}), \bibinfo{note}{unpublished}.

\bibitem[{\citenamefont{Starykh et~al.}(2000)\citenamefont{Starykh, Jacquod,
  Narimanov, and Stone}}]{SJNS00}
\bibinfo{author}{\bibfnamefont{O.~A.} \bibnamefont{Starykh}},
  \bibinfo{author}{\bibfnamefont{P.~R.~J.} \bibnamefont{Jacquod}},
  \bibinfo{author}{\bibfnamefont{E.~E.} \bibnamefont{Narimanov}},
  \bibnamefont{and} \bibinfo{author}{\bibfnamefont{A.~D.} \bibnamefont{Stone}},
  \bibinfo{journal}{Phys. Rev. E} \textbf{\bibinfo{volume}{62}},
  \bibinfo{pages}{2078} (\bibinfo{year}{2000}).

\bibitem[{\citenamefont{Bunimovich}(1974)}]{Bunimovich74b}
\bibinfo{author}{\bibfnamefont{L.~A.} \bibnamefont{Bunimovich}},
  \bibinfo{journal}{Funct. Anal. Appl.} \textbf{\bibinfo{volume}{8}},
  \bibinfo{pages}{254} (\bibinfo{year}{1974}).

\bibitem[{\citenamefont{Bunimovich}(1979)}]{Bunimovich79}
\bibinfo{author}{\bibfnamefont{L.~A.} \bibnamefont{Bunimovich}},
  \bibinfo{journal}{Commun. Math. Phys.} \textbf{\bibinfo{volume}{65}},
  \bibinfo{pages}{295} (\bibinfo{year}{1979}).

\bibitem[{\citenamefont{Benettin and Strelcyn}(1978)}]{BenS78}
\bibinfo{author}{\bibfnamefont{G.}~\bibnamefont{Benettin}} \bibnamefont{and}
  \bibinfo{author}{\bibfnamefont{J.-M.} \bibnamefont{Strelcyn}},
  \bibinfo{journal}{Phys. Rev. A} \textbf{\bibinfo{volume}{17}},
  \bibinfo{pages}{773} (\bibinfo{year}{1978}).

\bibitem[{\citenamefont{Shinohara et~al.}(2006)\citenamefont{Shinohara,
  Harayama, Tureci, and Stone}}]{SHTS06}
\bibinfo{author}{\bibfnamefont{S.}~\bibnamefont{Shinohara}},
  \bibinfo{author}{\bibfnamefont{T.}~\bibnamefont{Harayama}},
  \bibinfo{author}{\bibfnamefont{H.~E.} \bibnamefont{Tureci}},
  \bibnamefont{and} \bibinfo{author}{\bibfnamefont{A.~D.} \bibnamefont{Stone}},
  \bibinfo{journal}{Phys. Rev. A} \textbf{\bibinfo{volume}{74}},
  \bibinfo{pages}{033820} (\bibinfo{year}{2006}).

\bibitem[{\citenamefont{Fukushima et~al.}(2006)\citenamefont{Fukushima,
  Harayama, and Wiersig}}]{FHW06}
\bibinfo{author}{\bibfnamefont{T.}~\bibnamefont{Fukushima}},
  \bibinfo{author}{\bibfnamefont{T.}~\bibnamefont{Harayama}}, \bibnamefont{and}
  \bibinfo{author}{\bibfnamefont{J.}~\bibnamefont{Wiersig}},
  \bibinfo{journal}{Phys. Rev. A} \textbf{\bibinfo{volume}{73}},
  \bibinfo{pages}{023816} (\bibinfo{year}{2006}).

\bibitem[{\citenamefont{Lebental et~al.}(2007)\citenamefont{Lebental, Lauret,
  Zyss, Schmit, and Bogomolny}}]{LLZSB07}
\bibinfo{author}{\bibfnamefont{M.}~\bibnamefont{Lebental}},
  \bibinfo{author}{\bibfnamefont{J.~S.} \bibnamefont{Lauret}},
  \bibinfo{author}{\bibfnamefont{J.}~\bibnamefont{Zyss}},
  \bibinfo{author}{\bibfnamefont{C.}~\bibnamefont{Schmit}}, \bibnamefont{and}
  \bibinfo{author}{\bibfnamefont{E.}~\bibnamefont{Bogomolny}},
  \bibinfo{journal}{Phys. Rev. A} \textbf{\bibinfo{volume}{75}},
  \bibinfo{pages}{033806} (\bibinfo{year}{2007}).

\bibitem[{\citenamefont{Shinohara and Harayama}(2007)}]{SH07}
\bibinfo{author}{\bibfnamefont{S.}~\bibnamefont{Shinohara}} \bibnamefont{and}
  \bibinfo{author}{\bibfnamefont{T.}~\bibnamefont{Harayama}},
  \bibinfo{journal}{Phys. Rev. E} \textbf{\bibinfo{volume}{75}},
  \bibinfo{pages}{036216} (\bibinfo{year}{2007}).

\bibitem[{\citenamefont{Shinohara et~al.}(2007)\citenamefont{Shinohara,
  Fukushima, and Harayama}}]{SFH07}
\bibinfo{author}{\bibfnamefont{S.}~\bibnamefont{Shinohara}},
  \bibinfo{author}{\bibfnamefont{T.}~\bibnamefont{Fukushima}},
  \bibnamefont{and} \bibinfo{author}{\bibfnamefont{T.}~\bibnamefont{Harayama}},
  \bibinfo{journal}{arXiv:physics/07060106}  (\bibinfo{year}{2007}).

\bibitem[{\citenamefont{Harayama and Fukushima}(2004)}]{HF04}
\bibinfo{author}{\bibfnamefont{T.}~\bibnamefont{Harayama}} \bibnamefont{and}
  \bibinfo{author}{\bibfnamefont{T.}~\bibnamefont{Fukushima}},
  \bibinfo{journal}{J. Select. Top. Quantum Elec.}
  \textbf{\bibinfo{volume}{10}}, \bibinfo{pages}{1039} (\bibinfo{year}{2004}).

\bibitem[{\citenamefont{Fang et~al.}(2007)\citenamefont{Fang, Cao, and
  Solomon}}]{FYS07}
\bibinfo{author}{\bibfnamefont{W.}~\bibnamefont{Fang}},
  \bibinfo{author}{\bibfnamefont{H.}~\bibnamefont{Cao}}, \bibnamefont{and}
  \bibinfo{author}{\bibfnamefont{G.~S.} \bibnamefont{Solomon}},
  \bibinfo{journal}{Appl. Phys. Lett.} \textbf{\bibinfo{volume}{90}},
  \bibinfo{pages}{081108} (\bibinfo{year}{2007}).

\bibitem[{\citenamefont{Schwefel et~al.}(2004)\citenamefont{Schwefel, Rex,
  Tureci, Chang, and Stone}}]{SRTCS04}
\bibinfo{author}{\bibfnamefont{H.~G.~L.} \bibnamefont{Schwefel}},
  \bibinfo{author}{\bibfnamefont{N.~B.} \bibnamefont{Rex}},
  \bibinfo{author}{\bibfnamefont{H.~E.} \bibnamefont{Tureci}},
  \bibinfo{author}{\bibfnamefont{R.~K.} \bibnamefont{Chang}}, \bibnamefont{and}
  \bibinfo{author}{\bibfnamefont{A.~D.} \bibnamefont{Stone}},
  \bibinfo{journal}{J. Opt. Soc. Am. B} \textbf{\bibinfo{volume}{21}},
  \bibinfo{pages}{923} (\bibinfo{year}{2004}).

\bibitem[{\citenamefont{Lee et~al.}(2005)\citenamefont{Lee, Ryu, Kwon, Rim, and
  Kim}}]{LRKRK05}
\bibinfo{author}{\bibfnamefont{S.-Y.} \bibnamefont{Lee}},
  \bibinfo{author}{\bibfnamefont{J.-W.} \bibnamefont{Ryu}},
  \bibinfo{author}{\bibfnamefont{T.-Y.} \bibnamefont{Kwon}},
  \bibinfo{author}{\bibfnamefont{S.}~\bibnamefont{Rim}}, \bibnamefont{and}
  \bibinfo{author}{\bibfnamefont{C.-M.} \bibnamefont{Kim}},
  \bibinfo{journal}{Phys. Rev. A} \textbf{\bibinfo{volume}{72}},
  \bibinfo{pages}{061801(R)} (\bibinfo{year}{2005}).

\bibitem[{\citenamefont{Lee et~al.}(2007)\citenamefont{Lee, Yang, Moon, Lee,
  An, Shim, Lee, and Kim}}]{LYMLASLK07}
\bibinfo{author}{\bibfnamefont{S.-B.} \bibnamefont{Lee}},
  \bibinfo{author}{\bibfnamefont{J.}~\bibnamefont{Yang}},
  \bibinfo{author}{\bibfnamefont{S.}~\bibnamefont{Moon}},
  \bibinfo{author}{\bibfnamefont{J.-H.} \bibnamefont{Lee}},
  \bibinfo{author}{\bibfnamefont{K.}~\bibnamefont{An}},
  \bibinfo{author}{\bibfnamefont{J.-B.} \bibnamefont{Shim}},
  \bibinfo{author}{\bibfnamefont{H.-W.} \bibnamefont{Lee}}, \bibnamefont{and}
  \bibinfo{author}{\bibfnamefont{S.~W.} \bibnamefont{Kim}},
  \bibinfo{journal}{Phys. Rev. A} \textbf{\bibinfo{volume}{75}},
  \bibinfo{pages}{011802(R)} (\bibinfo{year}{2007}).

\bibitem[{\citenamefont{Wiersig and Hentschel}(2007)}]{WH07}
\bibinfo{author}{\bibfnamefont{J.}~\bibnamefont{Wiersig}} \bibnamefont{and}
  \bibinfo{author}{\bibfnamefont{M.}~\bibnamefont{Hentschel}},
  \bibinfo{journal}{To appear in Phys. Rev. Lett.}  (\bibinfo{year}{2007}).

\bibitem[{\citenamefont{Beck and Schl{\"o}gl}(1993)}]{BS93}
\bibinfo{author}{\bibfnamefont{C.}~\bibnamefont{Beck}} \bibnamefont{and}
  \bibinfo{author}{\bibfnamefont{F.}~\bibnamefont{Schl{\"o}gl}},
  \emph{\bibinfo{title}{Thermodynamics of chaotic systems}}
  (\bibinfo{publisher}{Cambridge University Press},
  \bibinfo{address}{Cambridge}, \bibinfo{year}{1993}).

\bibitem[{\citenamefont{Wiersig}(2000)}]{Wiersig00}
\bibinfo{author}{\bibfnamefont{J.}~\bibnamefont{Wiersig}},
  \bibinfo{journal}{Phys. Rev. E} \textbf{\bibinfo{volume}{62}},
  \bibinfo{pages}{R21} (\bibinfo{year}{2000}).

\bibitem[{\citenamefont{Lebental et~al.}(2006)\citenamefont{Lebental, Lauret,
  Hierle, and Zyss}}]{LLHZ06}
\bibinfo{author}{\bibfnamefont{M.}~\bibnamefont{Lebental}},
  \bibinfo{author}{\bibfnamefont{J.~S.} \bibnamefont{Lauret}},
  \bibinfo{author}{\bibfnamefont{R.}~\bibnamefont{Hierle}}, \bibnamefont{and}
  \bibinfo{author}{\bibfnamefont{J.}~\bibnamefont{Zyss}},
  \bibinfo{journal}{Appl. Phys. Lett.} \textbf{\bibinfo{volume}{88}},
  \bibinfo{pages}{031108} (\bibinfo{year}{2006}).

\bibitem[{\citenamefont{Sunada et~al.}(2005)\citenamefont{Sunada, Harayama, and
  Ikeda}}]{SHI05}
\bibinfo{author}{\bibfnamefont{S.}~\bibnamefont{Sunada}},
  \bibinfo{author}{\bibfnamefont{T.}~\bibnamefont{Harayama}}, \bibnamefont{and}
  \bibinfo{author}{\bibfnamefont{K.~S.} \bibnamefont{Ikeda}},
  \bibinfo{journal}{Phys. Rev. E} \textbf{\bibinfo{volume}{71}},
  \bibinfo{pages}{046209} (\bibinfo{year}{2005}).

\end{thebibliography}

\end{document}